
%
%
%
\input phyzzx
\unnumberedchapters
%
%
%
\def\eqnalign#1{\eqname{#1}}
%
%

%

%
\def\refoutlw{\par\penalty-400\vskip\chapterskip
   \spacecheck\referenceminspace
   \ifreferenceopen \Closeout\referencewrite \referenceopenfalse \fi
   \line{\bf\hfil References\hfil}\vskip\headskip
   \input \jobname.refs
   }
%
%
\def\a{\alpha}
\def\b{\beta}
\def\d{\delta}
\def\ga{\gamma}
\def\Ga{\Gamma}
\def\gfive{\gamma_{5}}
\def\m{\mu}
\def\n{\nu}
\def\r{\rho}
\def\s{\sigma}
\def\ee{\epsilon}
\def\gmn{ g_{\m\n}}
\def\pmpn{p_{\m} p_{\n}}
\def\La{\Lambda}
\def\FF{F^{a}_{\m\n}F^{a\,\m\n}}
\def\DFDF{(D^2 F_{\m\n})^a (D^2 F^{\m\n})^a}
\def\cb{\bar c}

%
%

\def\VV{\scriptscriptstyle V}

\def\SYM{S_{\rm YM}}
\def\SGF{S_{\rm GF}}
\def\SHCD{S_2}
\def\SHCDD{S_2^D}
\def\cv{c_{\VV}}
\def\SFGF{S^f_{\rm GF}}

\def\idx{\int \! d^4\!x\>}
\def\iddx{\int\! d^D\!x \>}
\def\ff{f^2 (\partial^2/ \La^2)}
\def\pf{f (\partial^2/ \La^2)}
\def\f{f\left({p^2\over \La^2}\right)}
\def\fator{ {g^2\cv \over 16\pi^2}}
\def\tmunu{p^2 g_{\m\n}-p_{\m}p_{\n}}
%

\mathsurround=2pt
\rightline{FTUAM 94/27}
\rightline{NIKHEF-H 94/27}
\date{}

%
\REF\Piguet{O. Piguet and A. Rouet, Phys. Rep. {\bf 76} (1981) 1.}
\REF\tHooft{G. `t Hooft and V. Veltman, Nucl. Phys. {\bf 44B} (1972)
 189.}
\REF\HCDPV{A.A. Slavnov, Nucl. Phys. {\bf 31B} (1971) 301.
\nextline B.W. Lee and J. Zinn-Justin, Phys. Rev. {\bf 5D} (1972)
 3121. }
\REF\Slavnov{A.A. Slavnov, Theor. Math. Phys. {\bf 33} (1977) 977.}
\REF\Breitenlohner{P. Breitenlohner and D. Maison, Commun. Math. Phys.
{\bf 52} (1977) 11.}
\REF\CPM{C.P. Martin, Phys. Lett. {\bf 241B} (1990) 513.}
\REF\GMR{G. Giavarini, C.P. Martin and F. Ruiz Ruiz, Nucl. Phys.
{\bf 381B} (1992) 222.}
\REF\Faddeev{L.D. Faddeev and A.A. Slavnov, {\it Gauge fields,
introduction to quantum theory}, second edition (Benjamin, Reading
1990) .}
\REF\Nogo{C.P. Martin and F. Ruiz Ruiz, {\it Higher covariant derivative
Pauli-Villars regularization does not lead to a consistent QCD}, FTUAM
 94/9, NIKHEF-H 94/24.}
\REF\Day{M. Day, Nucl. Phys. {\bf 231B} (1983) 1.}
\REF\Slavdr{A.A. Slavnov, {\it Symmetry preserving regularization for
gauge and supergauge theories}, in {\it Superspace and Supergravity},
edited by S.W. Hawking and M. Rocek, p.177 (Cambridge University Press,
Cambridge 1981).}
\REF\Bogoliubov{N.N. Bogoliubov and D.V. Shirkov, {\it Introduction to
the theory of quantized fields} Wiley-Interscience, New York.}
\REF\Epstein{ H. Epstein and V. Glaser, Ann. Inst. Henri Poincar\'e
{\bf 29} (1973) 211.}
\REF\CS{ G. Leibbrandt and C.P. Martin, Nucl. Phys. {\bf 377B} (1992)
593.
\nextline G. Giavarini, C.P. Martin and F. Ruiz Ruiz, Phys. Lett. {\bf
314B} (1993) 324.
\nextline G. Leibbrandt and C.P. Martin, Nucl. Phys. {\bf 416B} (1994)
351.
\nextline G. Giavarini, C.P. Martin and F. Ruiz Ruiz, Phys. Lett.
{\bf 332B} (1994) 345.}
\REF\Collins{J.C. Collins, {\it Renormalization }, Cambridge University
Press. Cambridge 1987.}
\REF\Itzykson{C. Itzykson and J-B Zuber, {\it Quantum Field Theory}
 McGraw-Hill Book Company.}
%

\titlepage

\title{{\seventeenbf Higher covariant derivative regulators and
non-multiplicative renormalization}}

\author{ C. P. Martin}
\address{{\it Departamento de F\'\i sica Te\'orica,  C-XI,
              Universidad Aut\'onoma de Madrid \break
              Cantoblanco, Madrid 28049, Spain}}

\author{ F. Ruiz Ruiz}
\address{{\it NIKHEF-H, Postbus 41882, 1009 DB Amsterdam,
              The Netherlands}}

\vskip 2 true cm

\noindent
The renormalization algorithm based on regularization methods with two
regulators is analyzed by means of explicit computations. We show in
particular that regularization by higher covariant derivative terms
can be complemented with dimensional regularization to obtain a
consistent renormalized 4-dimensional Yang-Mills theory at the
one-loop level. This shows that hybrid regularization methods can be
applied not only to finite theories, like \eg\ Chern-Simons, but
also to divergent theories.
\endpage
\pagenumber=2

\noindent
The renormalizability of a non-abelian gauge theory is not tied to the
existence of a regularization method preserving its BRS symmetry, but
to the more fundamental fact of the BRS symmetry not being anomalous
[\Piguet]. Yet, having gauge invariant regularizations at our disposal
is very useful for practical purposes. Sadly enough there are not many
BRS invariant regularization methods available, dimensional
regularization [\tHooft] and the method of higher covariant
derivatives [\HCDPV,\Slavnov] being the two most prominent ones.

Dimensional regularization is the most popular choice to regularize
vector-like theories since the algebraic structure of these is not
altered by a change in the dimension of spacetime. Ward identities
hold almost automatically in this case. Unfortunately, when chiral
objects like ${\gfive}$ matrices or Levi-Civita pseudotensors occur,
an algebraically consistent dimensional regularization
[\Breitenlohner] is rather awkward, should it exist at all
[\CPM,\GMR]. The reason for this is that the properties of chiral
objects depend on the dimension of spacetime, which clashes somehow
with the ideas behind dimensional regularization.

The method of higher covariant derivatives [\HCDPV,\Slavnov] is a very
interesting attempt to regularize non-abelian gauge theories while
keeping the spacetime dimension at its physical value. Think for
example of Yang-Mills theory. By adding suitable gauge invariant terms
to the classical action --see eq. (4) below-- one transforms the
theory into a superrenormalizable gauge invariant theory, henceforth
called higher covariant derivative theory. In this way partial
regularization of Yang-Mills theory is achieved in a manifestly gauge
invariant fashion. To regularize the few diagrams in the higher
covariant derivative theory that are primitively divergent and at the
same time preserve gauge invariance, one must supplement an additional
gauge invariant regulator. Since these diagrams happen to be one-loop,
finding a gauge invariant regulator that renders them finite should in
principle be easier than addressing the regularization of the original
Yang-Mills theory. The proposal in ref. [\Slavnov] for this additional
regulator is a certain explicitly gauge invariant Pauli-Villars
regularization whose latest version is explained in ref. [\Faddeev].
{}From now on, we will call higher covariant derivative Pauli-Villars
regularization to this combination of higher covariant derivatives and
Pauli-Villars regulators [\Faddeev].

It has been recently shown [\Nogo] that the latter combination leads
to a Yang-Mills theory with the wrong beta function. The reason being
that there are {\it non-local} contributions to the vacuum
polarization tensor that modify the {\it non-local} gauge-invariant
part of the effective action. This no-go result renders higher
covariant derivative Pauli-Villars regularization rather useless as a
general regularization method.  Nonetheless, since the anomalous
contributions come from the Pauli-Villars determinants themselves
[\Nogo], it might be possible that some other regularization of the
higher covariant derivative theory led to a sensible Yang-Mills
theory.

The purpose of this letter is to show that if we furnish the higher
covariant derivative theory with dimensional regularization, one
obtains a regularization method leading to a sensible 4-dimensional
Yang-Mills theory at the one-loop level. We will call this
regularization method higher covariant derivative dimensional
regularization and see that it is free of the problems that occur for
higher covariant derivative Pauli-Villars regularization.

In this state of affairs, one may rightfully ask why not using
dimensional regularization from the very beginning. In so doing we
would save quite a lot of cumbersome algebra coming from the higher
covariant derivative terms. In what follows we provide reasons to
motivate the regularization method under consideration. Let us first
consider perturbative three-dimensional Chern-Simons theory.  The
simplest higher covariant derivative term that one can add to the
Chern-Simons action is $F^2/\La,$ where $\La$ is a mass. This results
into a very interesting theory on its own, namely topologically
massive Yang-Mills theory, that when dimensionally regularized allows
explicit two-loop computations. These computations have been performed
in ref. [\GMR] and have shown the vanishing of the two-loop correction
to the bare Chern-Simons parameter. They have also proved that
topologically massive Yang-Mills theory is finite to {\it all} orders
in perturbation theory, even though by power counting the theory is
only superrenormalizable. With en eye on four-dimensional gauge
theories involving $\gfive$ and the Levi-Civita symbol
$\varepsilon^{\m\n\r\s},$ it is therefore important to see whether the
regularization method in question may become a general regularization
procedure or it only works in three dimensions. Another reason is that
if one shows that higher covariant derivative dimensional
regularization is consistent, one then has an argument to not
relinquish the possibility of a complete 4-dimensional regularization
based on higher covariant derivatives. Last, but not least, is that by
carrying on with this regularization method we shall learn how to
renormalize theories whose divergences are parametrized by two
regulators. This problem has not been tackled yet in a satisfactory
way in the existing literature [\Faddeev,\Day] even at one loop.

Let us start by formulating higher covariant derivative dimensional
regularization for 4-dimensional Yang-Mills theory. The classical
Yang-Mills action for gauge group $SU(N)$ in a covariant gauge
$\partial\!A^a=b^a$ is given by
$$
S = \SYM + \SGF \>,
\eqn\class
$$
where
$$
\eqalign{
& \SYM = {1\over 4}\idx \FF\cropen{8pt}
& \SGF = \idx  \!\left[{\a\over 2}\,b^a b^a -b^a\partial A^a
        + \big(J^{a\m} - \partial^\m \bar c^a\big)
          \big(D_{\m} c\big)^a
        - {1\over 2}\,f^{abc}\, H^a c^b c^c\, \right] \> .\cr}
\eqn\clasdef
$$
The action $S$ is invariant under BRS transformations
$$
sA^a_{\m} = (D_{\m} c)^a \;\;\;\; sb^a = 0 \;\;\;\;
s{\bar c}^a = b^a \;\;\;\; sc^a=-{1\over 2} g f^{abc} c^b c^c \;\;\;\;
sJ^a_{\m} = 0\;\;\;\;sH^a = 0,
\eqn\BRS
$$
the BRS operator $s$ being nilpotent, $s^2=0.$ We have introduced the
external fields $J^a_{\m}$ and $H^a$ to keep track of the
renormalization properties of the composite operators $sA^a_{\m}$ and
$sc^a,$ respectively. To achieve regularization of the ill-defined
Feynman diagrams corresponding to the action $S$ in eq. \class, one
proceeds in two steps [\Slavdr]. First, one introduces higher
covariant derivartive terms and replaces the classical gauge-fixed
action $S$ with the action
$$
S_\La = \SYM + \SHCD +  \SFGF \>,
\eqn\hcdact
$$
where  $\SYM$ is as in eq. \clasdef\ and $\SHCD$ and $\SFGF$ are
given by
$$
\eqalignno{
 & \SHCD = {1\over 4 {\La}^4}\idx \DFDF  & \eqnalign{\Stwo}
                                                    \cr
 & \SFGF = \idx \!\left[{\a\over 2}\,b^a {1\over \ff}\, b^a
   - b^a\partial A^a + \big(J^{a\m}
   - \partial^\m \bar c^a\big) \big(D_{\m} c^a\big)
   - {1\over 2}\,f^{abc}\, H^a c^b c^c\, \right] \!.
                                      & \eqnalign{\noname}\cr}
$$
Here $\a$ is the gauge fixing parameter and $\pf$ is given in momentum
space by
$$
\f = 1 + {p^4\over \La^4} ~,
$$
so as to ensure locality and render all $\a\!$-dependent contributions
finite by power counting [\Nogo]. In our notation, $\a=0$ corresponds
to the Landau gauge and $\a=1$ to the Feynman gauge. The action
$S_{\La}$ in eq. \hcdact\ is invariant under the BRS transformations
in eq. \BRS\ and gives rise to a superrenormalizable theory that we
shall call higher covariant derivative theory. The free gauge
propagator for this theory has the form
$$
G^{ab}_{\m\n}(p) = \d^{ab}\,
    {\La^4\over {p^4(p^4+\La^4)}}\, \left[\, \big( p^2 \gmn -
    \pmpn \big) + \a\, {\La^4\over p^4 + \La^4}\;\pmpn\,\right]\> .
$$
The ultraviolet behaviour of $G^{ab}_{\m\n}(p)$ is not strong
enough as to make finite by power counting all 1PI Feynman diagrams.
However, it is very easy to see that only one-loop 1PI diagrams
contributing to the two, three and four-point Green functions of the
gauge field are superficially divergent. In other words, the higher
covariant derivative theory defined by $S_\La$ is superrenormalizable.
To regularize it, we use dimensionally regularization and thus achieve
full regularization of the original Yang-Mills theory while preserving
BRS invariance.

Having set a manifestly BRS invariant regularization method, however
important this might be, is not the end of the story. Ultraviolet
divergences have to be subtracted if a renormalized theory is to be
obtained. In the case of gauge theories this means that given a
regularization method, an explicitly BRS invariant subtraction
procedure must be devised. Recall in this regard that dimensional
regularization and ordinary abelian Pauli-Villars regularization lead
to well established Bogoliubov-type subtraction algorithms
[\Breitenlohner,\Bogoliubov]. To the best of our knowledge, no
subtraction procedure has been shown to work for any higher covariant
derivative regularization of non-abelian gauge theories. It would
appear that setting such a procedure at one loop is a very simple
task. After all, if a regularization method is BRS invariant, removing
from the one-loop Green functions those contributions which are
stricitly UV divergent yields finite renormalized Green functions that
verify the BRS identities at one loop. This being true, one must not
forget that is not enough. General renormalization theorems in Lorentz
covariant perturbation theory [\Epstein] imply that renormalized 1PI
Green functions at one loop are unique modulo a polynomial in their
momenta. Hence, if a regularization process modifies the {\it
non-local} structure of the theory in the large cut-off limit, serious
dificulties will arise.  This is precisely what happens with higher
covariant derivative Pauli-Villars regularization.  In this case,
renormalization at one loop can be accomplished in a BRS invariant
manner and yet {\it non-local} gauge invariant radiative corrections
get modified so as to yield an unphysical beta function [\Nogo]. Let
us see that everything goes smoothly when higher covariant derivative
dimensional regularization is used.

Let $\Phi$ collectively denote the fields $A^a_{\m},\, c^a,\, \cb^a,\,
b^a,\, J^a_{\m}$ and $H^a,$ and let $\ee$ denote the dimensional
regulator defined by $D=4+2\ee$, $D$ being the spacetime dimension.
The regularized one-loop effective action $\Ga_{1}(\Phi,\ee,\La)$
obtained with the help of higher covariant derivative dimensional
regularization depends on two regulators, $\ee$ and $\La$. We want to
fully characterize its UV divergent behaviour. To do this, we first
report the one-loop values for the three 1PI functions that will allow
us to do so. The vacuum polarization tensor
$\Pi^{ab}_{\m\n}(p,\ee,\La),$ the ghost self-energy
$\Omega^{ab}(p,\ee,\La)$ and ghost vertex $V_{\m}^{ab}(k,p,\ee,\La)$
as computed with higher covariant derivatives and dimensional
regularization are given by
$$
\eqalign{
\Pi^{ab}_{\m\n}(p,\ee,&\La) =-\,\d^{ab}\big(\tmunu\big) \,
          {\scriptstyle \times} \cropen{8pt}
& {\scriptstyle \times} \bigg\lgroup \! 1 + \fator\>\bigg\{
    \!\! - {71\over 6}\, \bigg[ \, {1\over\ee}
         - \, \ln\!\bigg({\kappa^2\over\La^2}\bigg) \bigg]
         + \bigg({13\over 6}-{\a\over 2}\bigg)\,
           \ln\!\bigg({p^2\over\La^2}\bigg) + c_0(\a) \bigg\} \!\!
    \bigg\rgroup + \, {\rm vt}  \cr}
\eqn\vacpol
$$
$$
\Omega^{ab}(p,\ee,\La) = \left\{1+\fator\left[\,{3-\a\over 4}\,
  \ln\!\left({p^2\over\La^2}\right) + \omega_0(\a) \,\right] \right\}
  \d^{ab} p^2 + \,{\rm vt}\qquad\qquad\qquad\qquad\,
\eqn\ghost
$$
$$
V_{\m}^{ab}(k,p,\ee,\La) = -\,i g f^{abc}\bigg\{\bigg[ \, 1
  - \, \fator \, {\a\over 2} \, \ln\!\bigg({p^2\over \La^2}\bigg)
    \bigg] p_\m + \fator \, \a \,V_{\m}^{\rm fin}(k,p,\a) \bigg\}
  +\, {\rm vt} ~.
\eqn\ghostv
$$
Here ${\rm ``vt"}$ stands for contributions that vanish when one takes
the sequential limit $\ee\to 0,$ $\La\to\infty,$ $\kappa$ is the
dimensional regularization mass scale and $p^\m$ and $k^\m$ are
external momenta. The symbols $c_{0}(\a)$ and $\omega_0(\a)$ denote
constants which do not depend on $\kappa$ nor $\La.$ The function
$V_{\m}^{\rm fin}(k,p,\a)$ is finite, has mass dimension 1 and does
not depend on any of the dimensionful parameters $\kappa$ and $\La.$ A
number of comments regarding these expressions are in order.

\noindent
{\it Comment 1.} To obtain eqs. \vacpol-\ghostv, we have first
performed a Laurent expansion around $\ee = 0,$ then dropped the terms
that vanish as $\ee\to 0$ and finally discarded the terms that vanish
as $\La\to\infty.$ To do all this we have used the techniques
developed in ref. [\GMR]. Notice that this way to proceed implicitly
assumes that to obtain renormalized 1PI Green functions the regulators
$\ee$ and $\La$ should be removed in this order: first $\ee\to 0$ and
then $\La\to\infty.$ This prescription for removing the regulators is
rather natural for the regularization method we are considering and
the only well-defined one for Chern-Simons theory [\GMR,\CS].
Proceeding the other way around,
\ie\ taking first $\La\to\infty$ and then $\ee\to 0,$ is rather messy
and would eventually require renormalizing a theory at fixed but
arbitrary $\ee,$ something which is not clear how to achieve.

\noindent
{\it Comment 2.} The vacuum polarization tensor in eq. \vacpol\ has
two types of UV singularities, a simple pole at $\ee = 0$ and a
logarithmic divergence in $\La$. The pole comes from the fact that the
higher covariant derivative theory has a divergent propagator at one
loop. The singularity in $\La$ stems from the divergent structure of
Yang-Mills theory. Note in contrast that the ghost self-energy and the
ghost vertex do not show a singular behaviour at $\ee = 0.$ This is
due to the the fact that for finite $\La$ these two Green functions
are convergent by power counting at $D=4.$ As regards other Green
functions, it is not difficult to see using general results for
one-loop dimensionally regularized integrals [\Collins] and the
techniques in ref. [\GMR] that
\vskip -4pt
\item{(i)} only the three and four-point Green functions for the gauge
field will become singular at $\ee=0,$ the singularity being a single
pole, and
\vskip -4pt
\item{(ii)} that any other superficially divergent one-loop 1PI Green
function will develop UV divergent singularities in $\ln\La.$

\noindent
{\it Comment 3.} Using the identity
$$
\ln \big(p^2/\La^2\big) = \ln\big(p^2/M^2\big)
                        + \ln\big(M^2/\La^2\big) ~,
\eqn\trivial
$$
where $M$ is a mass scale (later on to become the renormalization
mass scale), we see that UV divergent contributions in the Green
functions \vacpol-\ghostv\ are polynomial in the external momenta.
Note that the non-local contributions explicitly displayed in eqs.
\vacpol-\ghostv\ are the same as those obtained with only dimensional
regularization. Furthermore, the techniques in ref. [\GMR] ensure that
also the non-local part in $V^{\rm fin}_\m(k,p,\a)$ in eq. \ghostv\ is
the same as for dimensional regularization, a fact that will be used
later.

We now turn to the computation of the UV divergent contribution to the
one-loop 1PI functional $\Ga_1 (\Phi,\ee,\La)$. Recalling that the
regularization method is explicitly BRS invariant and following
standard techniques [\Itzykson], we obtain that the one-loop
effective action $\Ga_1(\Phi,\ee,\La)$ satisfies the BRS equation
$$
\Delta_{\La,D}\,\Ga_1 (\Phi,\ee,\La) = 0\>,
\eqn\BRSLa
$$
where $\Delta_{\La,D}$ is the linearized Slavnov-Taylor operator for
the $D\!$-dimensional counterpart of the action $S_{\La}$ in eq.
\hcdact. The operator $\Delta_{\La,D}$ can be written as
$$
\Delta_{\La,D} = \Delta_{D}
   + \iddx {\d\SHCDD\over \d A^a_{\m}} \, {\d\over\d J^{a\m}}\>,
$$
where $\SHCDD$ is the $D\!$-dimensional version of the higher
covariant derivative action in eq. \Stwo\ and
$$
\Delta_D = \iddx \! \left[
    {\d S^D\over \d A^a_{\m}}\,{\d\over\d J^{a\m}} +
    {\d S^D\over \d J^a_{\m}}\,{\d\over\d A^{a\m}} +
    {\d S^D\over \d c^a}\,{\d\over\d H^a} +
    {\d S^D\over \d H^a}\,{\d\over\d c^a} +
    b^a{\d S^D\over\delta {\cb}^a}\,\right]
$$
is the Slavnov-Taylor operator for the gauge-fixed Yang-Mills action
in $D$ dimensions $S^D.$ In the limit $\ee\to 0,\,\La\to\infty$ we may
express $\Ga_1(\Phi,\ee,\La)$ as
$$
\Ga_1(\Phi,\ee,\La) = \Ga_1^{\rm div} (\Phi,\ee,\La)
                    + \Ga_1^{\rm fin} (\Phi) + {\rm ``vt"} \>,
$$
where again ${\rm ``vt"}$ denotes contributions that vanish as $\ee\to
0,\,\La\to\infty$. All finite contributions in this limit are
collected in $\Ga_1^{\rm fin}(\Phi),$ whereas $\Ga_1^{\rm div}
(\Phi,\ee,\La)$ only contains contributions which are divergent.  The
functional $\Ga_1^{\rm div}(\Phi,\ee,\La)$ is a four-dimensional
integrated local functional of the fields and its derivatives with
mass dimension 4 and ghost number 0. Its coefficients are of the form
$a_1\ee^{-1} + a_2\ln\La + a_3\ee^{-1}\ln\La,$  with $a_1,\, a_2$ and
$a_3$ constants. Using the methods in ref. [\GMR], it can actually be
shown that $a_3=0.$

Let us look at eq. \BRSLa. Since any one-loop 1PI diagram with at
least one $J_{\m}^{a}$ insertion is finite by power counting at $D=4,$
$J\!$-dependent contributions to $\Ga_1(\Phi,\ee,\La)$ are
finite as $\ee\to 0.$ Recalling that all superficially divegent 1PI
Green functions diverge as $\ln\La$ for $\La\to\infty$ and noting that
the action $\SHCD$ has an overall factor $1/\La^4,$ we have
$$
\lim_{\La\to\infty}\,\lim_{\ee\to 0}\;\;\iddx
   {\d\SHCDD \over \d A^a_{\m}} \, {\d\Ga_1\over\d J^{a\m}}\> = 0.
$$
Taking then the limit $\ee\to 0,\,\La\to\infty$ in eq. \BRSLa, it
follows that $\Ga_1^{\rm div}(\Phi,\ee,\La)$ satisfies
$$
\Delta\,\Ga_1^{\rm div} (\Phi,\ee,\La) = 0
$$
where $\Delta$ is the Slavnov-Taylor operator for the gauge-fixed
Yang-Mills action $S$ in 4 dimensions. The most general solution to
this equation has long been known [\Itzykson]; it has the form
$$
\Ga_1^{\rm div} (\Phi,\ee,\La) =  c_1\, \SYM + \Delta X\>,
\eqn\BRSsol
$$
where the functional $X$ is given by
$$
X= \idx \Big[ c_2 \big(J_{\m}^a -\partial_{\m}{\cb}^a\big)A^{a\m}
            + c_3 \,H^a c^a \Big]
\eqn\BRSsolX
$$
and $c_1,\,c_2$ and $c_3$ are coefficients. The coefficients
$c_1,\,c_2$ and $c_3$ can be obtained from the singular contributions
to three suitable 1PI Green functions. Choosing for these the vacuum
polarization tensor, the ghost self-energy and the ghost vertex in
eqs. \vacpol-\ghostv\ and using the identity \trivial, we find that
$$
\eqalign{
& c_1= \fator\,\bigg\{\! -\, {71\over 6}\,\bigg[\, {1\over\ee} -
       \ln\!\bigg({\kappa^2\over\La^2}\bigg)\bigg]
     + {11\over 3} \ln\!\bigg({M^2 \over \La^2}\bigg) \bigg\}
  \cropen{8pt}
& c_2= \fator ~ {3 + \a\over 4} ~ \ln\!\bigg({M^2 \over \La^2}\bigg)
  \cropen{8pt}
& c_3= \fator ~ {\a\over 2} ~ \ln\!\bigg({M^2\over\La^2}\bigg) ~. \cr}
\eqn\divcons
$$
Putting everything together, we have that $\Ga_1^{\rm
div}(\Phi,\ee,\La)$ is given by eqs. \BRSsol-\BRSsolX\ with
coefficients as in eq. \divcons.

Once we have characterized the UV divergent behaviour of
$\Ga_1(\Phi,\ee,\La),$ we are ready to define a renormalized BRS
invariant 1PI functional $\Ga_1^{\rm ren}(\Phi,g,M)$ that describes
Yang-Mills theory at the one-loop level. We define
$$
\Ga_1^{\rm ren}(\Phi,g,M)\equiv \lim_{\La\to\infty}\,\lim_{\ee\to 0}
    ~ (1-{\bf T})\> \Ga_1(\Phi,\ee,\La) ~,
\eqn\renac
$$
where
$$
{\bf T} \, \Ga_1(\Phi,\ee,\La)\equiv (c_1+f_1)\,S^D_{\rm YM}
                              + \Delta_D X^D \>,
$$
the functional $X^D$ reads
$$
X^D = \iddx  \Big[ (c_2 + f_2)
                       \big(J_\m^a - \partial_\m{\cb}^a \big) A^{a\m}
                     + (c_3 + f_3)\, H^a c^a \Big]
$$
and $f_1,\,f_2$ and $f_3$ are arbitrary $(\ee,\!\La)\!$-independent
constants.  Note that eqs. \BRSsol-\divcons\ guarantee that the
limit in eq. \renac\ is finite. Taking into account our considerations
above, it is very easy to realize that the equations
$$
\Delta_D\, {\bf T} \Ga_1(\Phi,\ee,\La)= 0 \qquad \qquad
   \lim_{\La\to\infty}\,\lim_{\ee\to 0}\;\;
   \iddx {\d S^D_2\over\d A^a_\m} \, {\d\over\d J^{a\m}}\,
   {\bf T}\Ga_1(\Phi,\ee,\La) = 0
$$
hold, thus having
$$
\lim_{\La\to\infty}\,\lim_{\ee\to 0} ~
    \Delta_{\La,D}\Ga_1^{\rm ren}(\Phi,g,M)
         = \Delta\Ga_1^{\rm ren}(\Phi,g,M)
$$
and
$$
\Delta\Ga_1^{\rm ren} (\Phi,g,M)= 0 \>.
$$
In other words, the renormalized one-loop functional defined in eq.
\renac\ is BRS invariant. In fact, this is the most general
1PI functional that can be constructed from the regularized functional
$\Ga_1(\Phi,\ee,\La)$ by performing local subtractions consistent with
BRS invariance. This is due to the well-known fact
[\Piguet,\Itzykson] that the most general finite renormalization
compatible with BRS invariance amounts to adding to the renormalized
1PI functional the quantity $\,f_1\,S+\Delta\idx\big[ f_2 (J
-\partial\cb)\,A + f_3 Hc \big],$ where $f_1,\,f_2$ and $f_3$ are
arbitrary finite constants. But this is already taken into account in
the definition of the Bogoliubov-type operator ${\bf T}.$

We now come to the computation of the one-loop beta function of the
renormalized theory. It is plain that the renomalized 1PI functional
can not be obtained by multiplicative renormalizations of the coupling
constant, the gauge parameter and the fields of the action $S_{\La}$
in eq.  \hcdact.  The reason is that such a multiplicative
renormalization will give rise to new divergences at $\ee=0$ since the
higher covariant derivative term $\SHCD$ will introduce divergent
counterterms not associated with any primitively divergent diagram.
Note that if we set $\a$ to zero and allow for a renormalization of
$\La,$ or equivalently multiply $\SHCD$ by a bare parameter
$\lambda_0,$ our renormalized 1PI functional can be obtained by
multiplycative renormalization of a regularized bare theory having
$S_{\La}$ as classical action [\Day].  Strictly speaking, this
renormalization procedure demands three renormalization conditions to
define Yang-Mills theory, for one has to define what is meant by
renormalized $\La,$ or equivalently renormalized $\lambda_0$.  It is
well-known [\Piguet] however that the construction of Yang-Mills
theory along the lines of renormalized perturbation theory [\Epstein]
requires only two renormalization conditions, as is the case with
dimensional regularization.  Hence, if we did renormalize $\La,$ we
would have to understand Yang-Mills theory as the large-$\!\La$ limit
of a renormalized theory having $S_{\La}$ as classical action.  In
this paper we are not concerned with this approach --it fails
altogether for $\a\neq 0$ [\Day].

Having no multiplicative connection between the renormalized theory
and a regularized bare theory, standard text book techniques can not
be used to compute beta functions and anomalous dimensions. Yet this
should pose no problem, since the knowledge of the renormalized Green
functions is all that is needed to fully characterize the theory. Let
us recall in this regard that for the rigorously established BPHZ
subtraction procedure there is no bare theory [\Piguet], the
renormalized theory is obtained directly and everything, if liable of
a perturbative calculation, can be computed from renormalized
quantities. Since the beta function and the anomalous dimensions are
the coefficients in the renormalization group equations for the
renormalized Green functions, they can be determined from the latter
equations. In our case the renormalization group equation for a 1PI
Green function $G_R(p_e,g,\a,M)$ reads\foot{Note that the renormalized
1PI functional $\Ga_1^{\rm ren}(\Phi,g,M)$ is defined by a
Bogoliubov-type subtraction from a regularized 1PI functional
compatible with a quantum action principle. This ensures that the
renormalization group equation holds true for the renormalized Green
functions.}
$$
\left[ M\,{\partial\over\partial M} + \b(g)\,{\partial\over\partial g}
     + {1\over 2}\,\sum_{\Phi} \ga_{\Phi}(g)\, N_{\Phi}
     + \d(g)\, {\partial\over\partial\a} \,
\right] G_R(p_e,g,\a,M) = 0 \> .
\eqn\reneq
$$
The renormalized vacuum polarization tensor, ghost self-energy and
ghost vertex in the minimal scheme $f_1=f_2=f_3=0$ are given by
$$
\eqalign{
&\Pi_R{}^{ab}_{\m\n}(p) = - \bigg\{\! 1 + \fator \, \bigg[
    \bigg({13\over 6}-{\a \over 2}\bigg) \,
                 \ln\!\bigg({p^2\over M^2}\bigg)
    + c_0(\a)\,\bigg] \!\bigg\} \,\d^{ab} \big(\tmunu\big)
\cropen{8pt}
&\Omega_R^{ab}(p) = \left\{ \! 1+\fator\, \left[ \,
      {3-\a\over 4}\, \ln\!\left({p^2\over M^2}\right)
   + \omega_0(\a)\, \right]\right\} \, \d^{ab} p^2  \cropen{8pt}
&V_R{}_{\,\m}^{abc}(k,p) =
-i g f^{abc}\bigg\{\bigg[ 1 - \fator \, {\a\over 2}\,
      \ln\!\bigg({p^2\over M^2}\bigg)\bigg] p_{\m}
   + \fator\,\a \,V_{\m}^{\rm fin}(k,p,\a)\bigg \}\>. }
\eqn\renpi
$$
Substituting these expressions in eq. \reneq\ and expanding in powers
of $g$ the functions $\b(g),$ $\ga_{\Phi}(g)$ and $\d(g)$, one obtains
a linear system to be verified by the coefficients of such series.
Solving this system leads to
$$
\eqalign{
&\b(g)= - \,{11\over 3}\, {g^3\cv \over 16\pi^2} +{\it O}(g^4) \qquad
   \qquad\qquad\quad  \ga_A (g) = \left( {13\over 3}-\a \right) \fator
                     +{\it O}(g^4) \cr
&\ga_c (g) + \ga_{\cb}(g)= (3-\a)\,\fator + {\it O}(g^4)
  \qquad~ \d(g) = {\it O}(g^2) \>. }
$$
These results are in complete agreement with the beta function and the
anomalous dimensions computed within any consistent mass independent
renormalization scheme known as yet, say the MS scheme for dimensional
regularization.

Let us make two final remarks. First, note that the non-local part of
the renormalized Green functions in eqs. \renpi\ are the same as for
any renormalization scheme based on dimensional regularization. This,
together with the fact that the renormalized 1PI functional can be
constructed from the three Green functions in eqs.
\renpi, implies that the renormalized effective action $\Ga_1^{\rm
ren}(\Phi,g,M)$ we have constructed here and the one-loop
dimensionally renrormalized 1PI functional differ by finite
renormalizations of the fields and the parameteres in the theory.
Our second remark concerns the existence to all orders in perturbation
theory of a consistent BRS invariant renormalization algorithm based
on higher covariant derivative dimensional regularization. Although
this problem lies well outside the scope of this paper, let us say
that the subtraction procedure should take care of the singularities
at $\ee=0$ first and then remove the divergent large-$\!\La$
behaviour. The BRS invariance of the algorithm should come from BRS
invariance of the regularization method and the fact the the purely
divergent UV contributions are BRS invariant on their own. Although
we do not have a rigorous proof of these statements, we have partially
checked them for two-loop diagrams with overlapping divergences.

\smallskip
\noindent{\bf Acknowledgements:} FRR was supported by FOM, The
Netherlands. The authors also acknowledge partial support from
CICyT, Spain.

\smallskip

\refoutlw

\end